\documentclass[fleqn,twoside]{article}
\def\be{\begin{equation}}
\def\ee{\end{equation}}
\def\bi{\bibitem}
\begin{document}

\title{If Gauss-Bonnet interaction plays the role of dark energy}
\author{Abhik Kumar Sanyal$^1$}\footnotetext[1]{Relativity and Cosmology
Research Centre,\\
\indent Department of Physics, Jadavpur University,\\ \indent Calcutta - 700032, India.\\
\indent Electronic address:abhikkumar@gmail.com;~~sanyal\_ak@yahoo.com\\
}
\maketitle
\begin{center}
Dept. of Physics, Jangipur College, Murshidabad,
\noindent
West Bengal, India - 742213. \\

\end{center}
\noindent
\begin{abstract}
Gauss-Bonnet-scalar interaction has been found to play a crucial
role from the beginning till the late time of cosmological
evolution. A cosmological model has been constructed where the
Universe starts with exponential expansion but with infinite
deceleration, $q\rightarrow \infty$ and infinite equation of state
parameter, $w\rightarrow \infty$. During evolution it passes
through the stiff fluid era, $q=2, w=1$, the radiation dominated
era, $q=1, w=1/3$ and the matter dominated era, $q=1/2, w=0$.
Finally, deceleration halts, $q=0, w=-1/3$, and it then encounters
a transition to the accelerating phase. Asymptotically the
Universe reaches yet another inflationary phase $q\rightarrow -1,
w\rightarrow -1$. Such evolution is independent of the form of the
potential and the sign of the kinetic energy term ie., even a
noncanonical kinetic energy is unable to phantomize $(w<-1)$ the
model.
\end{abstract}

 \noindent
\section{Introduction}
In the recent years lot of observations have been carried out that
lead to a precise knowledge of the cosmological evolution.
Important cosmological observations like abundance of galaxy
clusters \cite{na}, statistics of large scale redshift surveys
\cite{na}, angular power spectrum of cosmic microwave background
radiation (CMBR) \cite{dn} and baryon oscillations \cite{bo}
suggest that the Universe is nearly flat and almost $73\%$ of the
matter density is in the form of dark energy \cite{ms}. Further,
the magnitude-redshift relation from standard candles such as type
Ia supernovae (SnIa) \cite{ar} indicates that the Universe has
recently entered a phase of accelerating expansion. Reconciling
all these astronomical observations it is now quite clear that the
so called dark energy is slowly varying with negative pressure
\cite{lp} having repulsive properties that can encounter the
attractive gravitational force and the corresponding equation of
state parameter $(w=p/\rho)$ is presently pretty close to minus
one ($-1$) \cite{am}. So we can now brief the knowledge of the
cosmological evolution that we have gathered so far from all the
astronomical observations. Soon after the beginning, the Universe
passed through a phase of exponential expansion (inflation) and
during evolution it went through the stiff fluid era - when
pressure balanced the energy density $(p=\rho,\;ie.,\;w=1)$.
Thereafter, it encountered the radiation dominated era
$(p=\rho/3,\;ie.,\;w=1/3)$ and finally the Universe entered the
pressureless dust era $(p=0,\;ie.,\;w=0)$. Presently, as already
mentioned, it is accelerating with the equation state parameter
$w$ pretty close to minus one ($-1$) \cite{am}. So far all the
attempts made to construct different dark energy models of the
Universe encompassing all the observable phenomena went in vain
\cite{lp}, \cite{ns}. An alternative approach to accommodate dark
energy is to modify the General Theory of Relativity by
considering additional curvature invariant terms such as
Gauss-Bonnet (GB) term. GB term arises naturally as the leading
order of the $\alpha'$ expansion of heterotic superstring theory,
where, $\alpha'$ is the inverse string tension \cite{dj}. Some
interesting results appear in the literature with GB interaction.
Avoidance of naked singularities in dilatonic brane world
scenarios and the problem of fine tuning with scalar fields and GB
interaction have been discussed in \cite{ne}. Further in string
induced gravity near initial singularity GB coupling with scalar
field has been found \cite{ia} important for the occurrence of
nonsingular cosmology. In addition there are also some recent
investigations \cite{ri} in the context of dark energy models. In
these works issues (scalar-GB gravity is equivalent classically to
so-caled modified GB gravity) like phantom cosmology with GB
correction, interplay between GB term and quintessence scalar,
experimental constraints on astronomical and cosmological
observations and cosmological models with scalar dependent GB
interaction have been addressed.
\par
\noindent Gauss-Bonnet term is a topological invariant one and so
to get some contribution in the four dimensional space-time it
requires dynamic dilatonic scalar coupling. In this work we shall
consider such coupling and investigate cosmological consequence.
\par
\noindent The paper has been organized as follows. In the
following section we write the action and the field equations. In
section 3 we explore a set of solutions with negative GB
interaction. It is found that such solutions require noncanonical
kinetic energy. In section 4 we present another set of solutions
for both positive and negative GB interacting terms. It has been
observed that a noncanonical kinetic energy evolves through to a
canonical one, without affecting the cosmic evolution.
\section{Action and the field equations}
We start with the following action containing Gauss-Bonnet
interaction \be S=\int
d^4x\sqrt{-g}[\frac{R}{2\kappa^2}+
\frac{\Lambda(\phi)}{8}G(R)-g(\phi)\phi,_{\mu}\phi'^{\mu}-V(\phi)],
\ee where,
$G(R)=R^2-4R_{\mu\nu}R^{\mu\nu}+R_{\mu\nu\rho\sigma}R^{\mu\nu\rho\sigma}$
is the Gauss-Bonnet term which is appearing in the action with a
coupling parameter $\Lambda(\phi)$. In the action there is yet
another coupling parameter viz., $g(\phi)$. For the spatially flat
Robertson-Walker space-time
\[ds^2=-dt^2+a(t)^2[dr^2+r^2 d\theta^2+r^2 sin^2\theta d\phi^2]\]
The field equations are \be 2\frac{\ddot a}{a}+\frac{\dot
a^2}{a^2}=-\kappa^2[g\dot\phi^2-V(\phi)+2\Lambda'\dot\phi\frac{\dot
a\ddot a}{a^2}+(\Lambda'\ddot\phi+\Lambda''\dot\phi^2)\frac{\dot
a^2}{a^2}]=-8\pi G p, \ee

\be 3\frac{\dot a^2}{a^2}=\kappa^2[g\dot
\phi^2+V(\phi)-3\Lambda'\dot{\phi} \frac{\dot{a}^3}{a^3}]=8\pi G
\rho, \ee where, $p$ and $\rho$ are the effective pressure and
density generated by the scalar field and the Gauss-Bonnet
interaction. In addition we have got the $\phi$ variation equation

\be 2g(\ddot\phi+3\frac{\dot
a}{a}\dot\phi+\frac{1}{2}\frac{g'}{g}\dot\phi^2+\frac{V'}{2g})=3\Lambda'\frac{\dot
a^2 \ddot a}{a^3} \ee which may not be considered to be an
independent equation, since it is derivable from the above two
equations (2) and (3). In the above, over-dot and dash ($\prime$)
stand for differentiations with respect to the proper time $t$ and
$\phi$ respectively. Now, we are to solve for $a, \phi, g(\phi),
V(\phi)$ and $\Lambda(\phi)$ in view of the above two field
equations (2) and (3), which requires three assumptions. The first
assumption that we make is \be \Lambda'\dot\phi= \lambda, \ee
where $\lambda$ is a constant, which is physically reasonable,
since it implies that the Gauss-Bonnet coupling parameter
$\Lambda(\phi(t)) = \lambda t$ grows in time, and as a result it
might contribute at the later epoch of cosmological evolution.
Such an assumption also mathematically simplifies the field
equations (2) and (3)considerably, which are,

\be 2\dot H+3H^2=-\kappa^2[g\dot \phi^2-V(\phi)+2\lambda H\dot
H+2\lambda H^3]=-\kappa^2 p, \ee

\be 3H^2=\kappa^2[g\dot \phi^2+V(\phi)-3\lambda H^3]=\kappa^2\rho,
\ee where, $H= \dot a/a$. Now eliminating $V(\phi)$ between
equations (6) and (7) we find, \be \dot H+\kappa^2[g\dot
\phi^2+\lambda H\dot H-\frac{\lambda}{2}H^3]=0. \ee One can also
eliminate $g\dot\phi^2$ between the same pair of equations to
express the potential as,

\be
V=\frac{3}{\kappa^2}H^2+\frac{5\lambda-n^2}{2}H^3-\frac{\kappa^2
n^2\lambda}{2}H^4. \ee Hence, we shall now deal with equations
(5), (7), (8) and/or (9) to solve the field variables $a, \phi$
the potential $V(\phi)$ and the parameters $g, \lambda$ of the
theory. Two important parameters that we deal with, in the
cosmological context, are the equation of state
($w=\frac{p}{\rho}$) and the deceleration ($q =-\frac{a\ddot
a}{\dot a^2}$) parameters, which are expressed as,

\be w=-1-\frac{2\dot H}{3H^2} \;\;\&\;\; q=-1-\frac{\dot
H}{H^2}=\frac{1+3w}{2}.\ee
\section{Solution with negative Gauss-Bonnet interaction}
The Gauss-Bonnet term interacts with the dilatonic scalar field
through $\Lambda(\phi)$ which may appear with both the sign
positive and negative. Here we consider the negative Gauss-Bonnet
interaction. Let us assume, \be g\dot\phi^2+\lambda H\dot H=0 \ee
Under the above choice, equation (8) gets solved immediately to
yield \be H=\frac{1}{\kappa n \sqrt
t}\;\;\&\;\;a=a_{0}\;e^{\frac{2\sqrt t}{\kappa n}},\ee where,
$\lambda=-n^2$ and we have considered positive sign only for both
$\kappa$ and $n$ to ensure expanding model. Now in view of
equation (10) we can find the state and the deceleration
parameters as \be w=-1+\frac{\kappa n}{3\sqrt
t}\;\;\&\;\;q=-1+\frac{\kappa n}{2\sqrt t}\;. \ee It is to be
noted that the above set of solutions (12) and (13) depends on the
constant $n$, which determines the GB coupling parameter
$\Lambda(\phi)$. Thus at the very early stage of the evolution of
the Universe, ie., at $t\rightarrow 0$, $q\rightarrow \infty$ and
$w\rightarrow \infty$. Thereafter at $t=\frac{\kappa^2 n^2}{36}$
the Universe enters stiff fluid era with $q=2$ and $w=1$.
Radiation dominated era starts at $t=\frac{\kappa^2 n^2}{16}$,
when $q=1$ and $w=\frac{1}{3}$. At $t=\frac{\kappa^2 n^2}{9}$ the
Universe becomes dust filled, with $q=\frac{1}{2}$ and $w=0$.
There is a transition from the decelerating to the accelerating
Universe at $t=\frac{\kappa^2 n^2}{4}$, when $q=0$ and
$w=-\frac{1}{3}$. Finally with the increase of the proper time
$q\rightarrow -1$ and $w\rightarrow -1$ asymptotically. It is most
important to notice that such evolution from the very early to the
late time of the Universe encompassing all the experimental
observations is independent of the form of the potential
$V(\phi)$.
\par
\noindent Now in view of equation (11) we get \be
g\dot\phi^2=-\frac{\kappa^2 n^2}{2}H^4=-\frac{1}{2\kappa^2 t^2}.
\ee Thus the solutions obtained demands $g(\phi)$ has to be
negative, ie the kinetic term appears with a wrong sign. It is
rather interesting to note that even a wrong sign of kinetic
energy does not phantomize \cite{p} the cosmological model under
consideration ie., the state parameter $w$ never goes beyond $-1$.
However, to find explicit solutions of the model under
consideration, we have to fix up either $g(\phi)$, $\Lambda(\phi)$
$\phi$ or the potential $V(\phi)$. As an example we consider the
most natural choice, viz.,  $g=-\frac{1}{2}$. Thus equation (14)
gets solve for $\phi$ as \be \phi=\frac{\ln t}{\kappa}.\ee
Equation (5) now solves $\Lambda (\phi)$ as \be \Lambda =-n^2
e^{\kappa\phi}.\ee and equation (7) solves the potential as \be
V=\frac{1}{2\kappa^4
n^2}[6e^{-\kappa\phi}+\kappa^2n^2e^{-2\kappa\phi}-6\kappa
ne^{-\frac{3\kappa\phi}{2}}].\ee It is to be mentioned that the GB
coupling parameter $\Lambda$ is expressed (eg., N.E.Mavromatos and
J.Rizos in \cite{ne}) in the form, \be
\Lambda=-\lambda_{0}e^{l\phi},\ee where, $l=-4/\sqrt 6$ in four
dimension. So, instead of fixing $g$ if one chooses $\lambda$ as
in (18), then the solutions are found as \be
\phi=\frac{1}{l}\ln{\frac{\lambda_{0}}{n^2
t}};\;\;\&\;\;g=-\frac{l^2}{2\kappa^2},\ee and \be
V=\frac{1}{2\kappa^4 \lambda_{0}^2}[2\lambda e^{l\phi}+n^4
e^{2l\phi}-3\kappa n^2 \sqrt{\lambda_{0}}e^{\frac{3l\phi}{2}}].
\ee It is to be noted that the GB coupling parameter
$\Lambda(\phi)$ and the potential $V(\phi)$ carry exponents with
opposite signs automatically, and asymptotically the potential
becomes a constant. As mentioned, one can also choose different
forms of the potential to find explicit solutions which we shall
not consider in this section. What we have observed is that the
negative interaction with Gauss-Bonnet term leads to a
noncanonical kinetic energy, which has got some interest in the
context of phantom cosmology \cite{p}, but not in general. In the
following section we study both positive and negative interactions
and transition from noncanonical to canonical kinetic energy.
\section{Solution with both signs of Gauss-Bonnet interaction}
In the previous section we have observed that the condition (11)
led to negative GB interaction which ultimately made the kinetic
energy noncanonical. In this section we make a different
assumption so that the GB interaction $\Lambda(\phi)$ may be
positive as well. Let us consider, \be g \dot\phi^2+\lambda H\dot
H -\frac{\lambda}{2}H^3=\frac{n_{1}^2}{2}H^3, \ee where, $n_{1}$
is a constant. As a result equation (8) is solved to yield \be
H=\frac{1}{\kappa n_{1} \sqrt t},\;\;\&\;\;a=a_{0}e^{\frac{2\sqrt
t}{\kappa n_{1}}}.\ee Further, equation (10) can be expressed as,
\be w=-1+\frac{\kappa n_{1}}{3\sqrt t}\;\;\&\;\;q=-1+\frac{\kappa
n_{1}}{2\sqrt t}.\ee Thus we obtain the same set of solutions as
in the previous section with the only difference that the constant
parameter $n$ that determines the solutions (12) and (13) in the
previous section 3, determines the GB coupling parameter
$\Lambda(\phi)$ too, while in the above solutions (22) and (23)
$n_{1}$ has nothing to do with $\Lambda(\phi)$. Thus $\lambda$
here may be positive as well as negative and so is the GB
interaction parameter $\Lambda(\phi)$.The Universe evolves in the
same manner as discussed in the previous section, starting from an
exponential expansion with infinite deceleration and passing
through the phases of stiff fluid, radiation dominated and the
matter dominated era. It then finally encounters a transition to
the accelerating phase when $w=-1/3$ and asymptotically reaches
$q=-1,w=-1$. We would like to mention once again that such
cosmological evolution remains independent of the form of
potential $V(\phi)$. Equation (21) now reduces to \be
g\dot\phi^2=\frac{1}{\kappa^2
n_{1}^2}[\frac{\lambda}{t^2}+\frac{\lambda+n_{1}^2}{\kappa n_{1}
t^{3/2}}].\ee To find the complete set of solutions we can explore
different situations fixing up either $g(\phi), \Lambda(\phi)$ or
the potential $V(\phi)$. In the following we cite a few examples.
\par
\noindent {\bf{Case I}}
\par
\noindent The most natural choice, $g= 1/2$ leads to a well
behaved solution, with \be \phi= \frac{4}{(\kappa
n_{1})^{3/2}}\left[\{\lambda \kappa n_{1}+(\lambda+n_{1}^2)\sqrt
t\}^{1/2}+\frac{\sqrt{\lambda\kappa
n_{1}}}{2}\ln\left\{\frac{\sqrt{\lambda\kappa
n_{1}+(\lambda+n_{1}^2)\sqrt t}-\sqrt{\lambda\kappa
n_{1}}}{\sqrt{\lambda\kappa n_{1}+(\lambda+n_{1}^2)\sqrt
t}+\sqrt{\lambda\kappa n_{1}}}\right\}\right],\ee while the
potential can be expressed as a function of time as, \be
V=\frac{1}{\kappa^2n_{1}^2}[\frac{3}{\kappa^2
t}+\frac{5\lambda-n_{1}^2}{2\kappa n_{1} t\sqrt
t}-\frac{\lambda}{2t^2}],\ee but it is extremely difficult to
express the potential and the GB interaction parameter $\Lambda$
as a function of $\phi$ .
\par
\noindent {\bf{Case II}}\par\noindent Next let us choose $\Lambda$
as, \be \Lambda=\frac{\beta}{\phi}.\ee Thus $\phi$ can be solved
in view of equation (5) as, \be \phi=\frac{\beta}{\lambda t},\ee
which decreases while $\Lambda$ increases linearly with time.
$g(\phi)$ can be found in view of equation (24) as, \be
g=\sqrt{\frac{\beta}{\lambda}}(\frac{n_{1}^2+\lambda}{\kappa^3
n_{1}^3})\phi^{-5/2}+\frac{\lambda}{2\kappa^2
n_{1}^2}\phi^{-2}.\ee Thus $g(\phi)\rightarrow \infty$
asymptotically. Now the potential takes the following form, \be
V=\frac{\lambda}{\kappa^2 n_{1}^2
\beta}[\frac{3}{\kappa^2}\phi+\sqrt{\frac{\lambda}{\beta}}
(\frac{5\lambda-n_{1}^2}{2\kappa
n_{1}})\phi^{3/2}-\frac{\lambda^2}{2\beta}\phi^2].\ee It is
interesting to observe that if one sets $\lambda=-m^2$, then
$\beta$ has to be negative making the Gauss-Bonnet interaction
parameter negative. For such a situation, with $\beta=-c^2$, we
have \be g=\frac{c}{m}(\frac{n_{1}^2-m^2}{2\kappa^3
n_{1}^3})\phi^{-5/2}-\frac{m^2}{2\kappa^2n_{1}^2}\phi^{-2}.\ee
Hence, for $n_{1}^2>m^2$, $g(\phi)<0$ at the beginning but becomes
positive at a later stage of the cosmic evolution. So a
noncanonical kinetic energy turns canonical at \be
\phi<\frac{c^2(n_{1}^2-m^2)^2}{\kappa^2n_{1}^2m^6},\;\;ie., at
\;\;t>\frac{\kappa^2n_{1}^2m^4}{(n_{1}^2-m^2)^2}.\ee This proper
time has nothing to do with the transitions of the state parameter
or the deceleration parameter in general. So even though a
noncanonical kinetic energy evolves through to a canonical one, it
does not play any role in the evolution of the Universe and the
state parameter $w$ always remains over the phantom divide line.
However, by properly choosing $m$ in terms of $n_{1}$, eg.,
choosing $m = n_{1}/\sqrt 3$, one can relate the time of flipping
of the sign of the kinetic energy term to the time of transition
of a decelerating Universe to an accelerating one.
\par
\noindent {\bf{Case III}}\par\noindent Let us now choose $\Lambda$
in the form \be \Lambda=\beta e^{l\phi}\ee where, $\beta$ is a
constant. So, in view of equations (5) and (24)\be
\phi=\frac{1}{l}\ln{(\frac{\lambda}{\beta}t)},\;\;\&\;\;g=\frac{l^2}{2\kappa^2
n_{1}^2}[\lambda+\sqrt{\frac{\beta}{\lambda}}(\frac{\lambda+n_{1}^2}{2\kappa
n_{1}})e^{\frac{l}{2}\phi}],\ee and the potential can be found
from equation (9) as, \be V=\frac{\lambda}{2\kappa^4 n_{1}^4
\beta^2}[6\beta n_{1}^2e^{-l\phi}-\kappa^2 n_{1}^2\lambda^2
e^{-2l\phi}+\kappa
n_{1}\sqrt{\beta\lambda}(5\lambda-n_{1}^2)e^{-\frac{3l\phi}{2}}].\ee
For, $l=-4/\sqrt 6$, $\phi$ becomes negative but that does not
create any problem whatsoever, and the potential carries positive
exponents. As before, here again we can choose $\lambda=-m^2,
\beta=-c^2$ and $l=-s^2$ and find \be
g=\frac{s^4}{2\kappa^2n_{1}^2}[-m^2+\frac{n_{1}^2-m^2}{2\kappa
n_{1}}(\frac{c}{m})e^{-\frac{s^2\phi}{2}}].\ee Now since
$\phi=\frac{1}{s^2}[\ln{(\frac{c^2}{m^2 t})}]$, so at
$t\rightarrow 0, \phi\rightarrow \infty$ and $g<0$. It has a
turning point at a later epoch $t=\frac{4\kappa^2 n_{1}^2
m^4}{(n_{1}^2-m^2)^2}$ after which $g$ becomes positive provided
$n_{1}>m$. So, here again we encounter the same situation as
discussed in case II, that the solutions remain unaffected even
though a noncanonical kinetic energy evolves through to a
canonical one.
\par
\noindent {\bf{Case IV}}\par\noindent Here we choose a simple
quadratic form of the potential, viz., \be
V(\phi)=\lambda_{0}\phi^2=\frac{V_{0}^2}{2\kappa^2}\phi^2.\ee As a
result, \be \phi=\frac{1}{\kappa n_{1} V_{0}}[\frac{6}{\sqrt
t}-\frac{\kappa(n_{1}^2-5\lambda)}{t}-\frac{\kappa^2\lambda}{t\sqrt
t}]^{1/2}.\ee Here, for a positive GB interaction
$\lambda>0,\Lambda>0$ even an imaginary scalar field evolves to a
real one, without affecting the nature of cosmic evolution. In
view of equations (5) and (24) it is now in principle possible to
find the forms of $\Lambda(\phi)$ and $g(\phi)$.
\section{Concluding remarks}
A viable cosmological model has been presented in four dimensions
considering Gauss-Bonnet-scalar coupling. To explain recent
accelerating expansion of the Universe the Gauss-Bonnet term
should dominate at the later epoch of cosmological evolution.
Therefore under such physically reasonable assumption that the
coupling parameter $(\Lambda)$ grows linearly in time a set of
solutions have been presented which are independent of the
signature of the coupling parameter $(\Lambda)$. The solutions
demonstrate that the Universe starts with an exponential
expansion, but with infinite deceleration $(q\rightarrow\infty)$
and the equation of state $(w\rightarrow\infty)$ parameters. In
the process of evolution it then passes through the stiff fluid
era $(w=1)$, the radiation dominated era $(w=1/3)$ and the matter
dominated (pressureless dust) era $(w=0)$. It then encounters a
turning point when ($w=-1/3)$, after which the Universe starts
accelerating. Asymptotically, both the deceleration and the
equation of state parameters go over to $-1$. Hence the
observations suggest that we are now living at the final stage of
cosmological evolution and the dark energy is presently evolving
rapidly from $w=-1/3$ to $w=-1$. For a negative coupling parameter
$(\Lambda<0)$, the kinetic energy is noncanonical. It has been
observed that a noncanonical kinetic energy might evolve to a
canonical one without influencing the nature of the solutions in
any way, and the equation of state parameter $(w)$ never goes
beyond the phantom divide line. Even an imaginary scalar field
might evolve to a real one without affecting the cosmological
evolution. Thus dark energy in the form of Gauss-Bonnet
interaction has been found to play a crucial role in the
cosmological evolution.

\end{document}